\documentclass[a4paper, twoside, fontsize=9pt, twocolumn]{scrartcl}
\pdfoutput=1

\usepackage[a4paper, top=88pt, bottom=88pt, left=50pt, right=50pt, headsep=16pt, footskip=28pt]{geometry}
\usepackage{amsfonts,amssymb,amsmath,amsthm,amstext,amssymb,amsopn,mathtools,nicefrac,xfrac}
\usepackage[authoryear,sort,round]{natbib} 
\usepackage[utf8]{inputenc}                 
\usepackage{booktabs}                       
\usepackage{nicefrac}                       
\usepackage{microtype}                      
\usepackage{graphics,graphicx}              
\usepackage{subcaption}                     
\usepackage{enumitem}
\usepackage{lipsum}  

\usepackage[boxruled,linesnumbered]{algorithm2e} 
\SetKwComment{Comment}{\%}{}
\SetKwInput{KwInput}{Input}
\SetKwInput{KwOutput}{Output}

\usepackage{verbatim}                       
\usepackage{appendix}                       
\usepackage{bm}                             
\usepackage{array}                          
\newcolumntype{C}[1]{>{\centering\arraybackslash}m{#1}}  
\usepackage[usenames,dvipsnames]{xcolor}    
\usepackage{hyperref}                       
\hypersetup{colorlinks=true,linkcolor=Maroon,filecolor=Magenta,urlcolor=Blue,citecolor=RoyalBlue}
\usepackage[noabbrev,capitalise,nosort,nameinlink]{cleveref}  
\usepackage{bbm}                            
\usepackage{mathtools}                      
\usepackage{todonotes}

\newcommand*{\arXiv}[1]{\bgroup\color{blue}\href{https://arxiv.org/abs/#1}{arXiv:#1}\egroup}
\newcommand*{\doi}[1]{\bgroup\color{blue}\href{https://doi.org/#1}{doi:#1}\egroup}
\newcommand*{\email}[1]{\bgroup\color{blue}\href{mailto:#1}{#1}\egroup}
\renewcommand*{\url}[1]{\bgroup\color{blue}\href{#1}{#1}\egroup}
\usepackage{enumitem, moreenum}
\setlist[enumerate]{nosep}
\setlist[itemize]{nosep}
\usepackage{mleftright} \mleftright
\renewcommand{\qedsymbol}{$\blacksquare$}
\renewenvironment{proof}[1][\proofname]{\noindent{\bfseries\sffamily #1.} }{\hfill\qedsymbol\medskip}
\usepackage[textfont={small}, labelfont={sf,bf,small},format=plain,indention=0cm]{caption}
\DeclareCaptionLabelSeparator{figlabelsep}{\,\,\,}
\usepackage{scrlayer-scrpage, xhfill}
\automark[section]{section}
\setkomafont{pageheadfoot}{\normalcolor\sffamily}
\setkomafont{pagenumber}{\normalfont\normalsize\sffamily}
\clearpairofpagestyles
\let\oldtitle\title
\renewcommand{\title}[1]{\oldtitle{#1}\newcommand{\theshorttitle}{#1}}
\newcommand{\shorttitle}[1]{\renewcommand{\theshorttitle}{#1}}
\let\oldauthor\author
\renewcommand{\author}[1]{\oldauthor{#1}\newcommand{\theshortauthor}{#1}}
\newcommand{\shortauthor}[1]{\renewcommand{\theshortauthor}{#1}}
\cohead{\xrfill[0.525ex]{0.6pt}~\theshorttitle~\xrfill[0.525ex]{0.6pt}}
\cehead{\xrfill[0.525ex]{0.6pt}~\theshortauthor~\xrfill[0.525ex]{0.6pt}}
\newcommand{\theabstract}[1]{\par\bgroup\noindent\textbf{\textsf{Abstract.}} #1\egroup}
\newcommand{\thekeywords}[1]{\par\smallskip\bgroup\noindent\textbf{\textsf{Keywords.}}\newcommand{\and}{ $\bullet$ } #1\egroup}
\newcommand{\themsc}[1]{\par\smallskip\bgroup\noindent\textbf{\textsf{2020 Mathematics Subject Classification.}}\newcommand{\and}{ $\bullet$ } #1\egroup}
\newcommand*{\affilref}[1]{\ref{affiliation#1}}
\newcommand*{\affiliation}[3]{
	\footnotetext[#1]{\label{affiliation#2} #3}
}
\usepackage{siunitx} 

\numberwithin{equation}{section}
\numberwithin{figure}{section}
\numberwithin{table}{section}

\newcommand*{\defeq}{\coloneqq}

\newcommand*{\Reals}{\mathbb{R}}

\newcommand*{\Natural}{\mathbb{N}}

\newcommand*{\posE}{\text{e}}

\theoremstyle{definition}

\crefname{assumption}{Assumption}{Assumptions}
\Crefname{assumption}{Assumption}{Assumptions}

\title{Multiple solutions to the static forward free--boundary Grad--Shafranov problem on MAST-U}
\shorttitle{Multiple free--boundary GS solutions on MAST-U}
\author{
    K.~Pentland\textsuperscript{\affilref{UKAEA}}
    \and
    N.~C.~Amorisco\textsuperscript{\affilref{UKAEA}} 
    \and
    P.~E.~Farrell\textsuperscript{\affilref{Oxford},\affilref{Prague}}
    \and
    C.~J.~Ham\textsuperscript{\affilref{UKAEA}}
}
\shortauthor{K.~Pentland, N.~C.~Amorisco, P.~E.~Farrell, and C.~J.~Ham}
\date{\today}

\begin{document}
\maketitle

\affiliation{1}{UKAEA}{United Kingdom Atomic Energy Authority, Culham Campus, Abingdon, Oxfordshire, OX14 3DB, United Kingdom\newline (\email{kamran.pentland@ukaea.uk})}
\affiliation{2}{Oxford}{Mathematical Institute, University of Oxford, Oxford, Oxfordshire, OX2 6GG, United Kingdom}
\affiliation{3}{Prague}{Mathematical Institute, Faculty of Mathematics and Physics, Charles University, Prague 186 75, Czechia}


    
\begin{abstract}\small
    \theabstract{%
    The Grad--Shafranov (GS) equation is a nonlinear elliptic partial differential equation that governs the ideal magnetohydrodynamic equilibrium of a tokamak plasma.
    Previous studies have demonstrated the existence of multiple solutions to the GS equation when solved in idealistic geometries with simplified plasma current density profiles and boundary conditions.
    Until now, the question of whether multiple equilibria might exist in real-world tokamak geometries with more complex current density profiles and integral free-boundary conditions (commonly used in production-level equilibrium codes) has remained unanswered.
    In this work, we discover multiple solutions to the static forward free-boundary GS problem in the MAST-U tokamak geometry using the validated evolutive equilibrium solver FreeGSNKE and the deflated continuation algorithm.
    By varying the plasma current, current density profile coefficients, or coil currents in the GS equation, we identify and characterise distinct equilibrium solutions, including both deeply and more shallowly confined plasma states.
    We suggest that the existence of even more equilibria is likely prohibited by the restrictive nature of the integral free-boundary condition, which globally couples poloidal fluxes on the computational boundary with those on the interior.
    We conclude by discussing the implications of these findings for wider equilibrium modelling and emphasise the need to explore whether multiple solutions are present in other equilibrium codes and tokamaks, as well as their potential impact on downstream simulations that rely on GS equilibria.
    }
    \thekeywords{%
        {Multiple solutions}%
        \and%
        {Grad--Shafranov}%
        \and%
        {MHD equilibria}%
        \and%
        {FreeGSNKE}%
        \and%
        {Deflated continuation}%
        \and%
        {MAST-U}%
    }
\end{abstract}
 

\section{Introduction} \label{sec:intro}

\subsection{Motivation and aims} 

The solution to the static forward free--boundary Grad--Shafranov (GS) problem describes the magnetohydrodynamic (MHD) equilibrium state of a magnetically-confined, toroidally symmetric plasma in a tokamak fusion device.
Obtaining accurate solutions to this problem is a critical requirement for experimental tokamak plasma design and operation.
The GS equation is a nonlinear elliptic partial differential equation (PDE) that can be solved in different tokamak geometries with a wide range of plasma current density parametrisations.
Due to its nonlinearity, the GS problem may support multiple, isolated solutions. 
However, computational methods used for the simulation or reconstruction of free--boundary GS equilibria almost always return only a single solution to the problem.
This solution is then used in a computational plasma simulation pipeline (e.g.\ in stability analysis, scenario modelling, and tokamak optimisation), without consideration of other possible solutions and how they might impact subsequent analyses.

In this paper, we build upon the work of \cite{ham2024}, which identified multiple numerical solutions to a contrived fixed-boundary GS problem. We demonstrate the existence of multiple solutions to the static forward (integral) free-boundary GS problem for real-world plasmas in the MAST-U tokamak.
We aim to:
\begin{enumerate}[label=(\roman*)]
    \item find multiple solutions to the static forward GS problem on MAST-U using \emph{FreeGSNKE} and the \emph{deflated continuation} algorithm;
    \item investigate how they change (and possibly bifurcate) when certain parameters in the GS equation are varied.
\end{enumerate}

The presence of multiple GS solutions in real-world tokamaks could have significant implications for a number of different areas across plasma simulation.
For example, this may also occur during the integrated modelling of plasma scenarios \citep{romanelli2014}, such as those on ITER or STEP \citep{chapman2024}, which require the time-dependent evolution of an equilibrium alongside coil currents and plasma profile parameters (e.g.\ using transport codes).
Missing bifurcations points and therefore the presence of multiple solutions during such a simulation could potentially undermine scenario design and operational planning.
These implications highlight the need to explore and identify different solution branches during forward/inverse equilibrium simulations.

To solve the forward GS problem here, we will make use of the Python-based, dynamic free--boundary toroidal plasma equilibrium solver FreeGSNKE, developed by \cite{amorisco2024}. 
This solver has the ability to carry out both (static/dynamic) forward and (static-only) inverse free--boundary GS equilibrium calculations. 
The static forward solver has previously been validated against the equilibrium codes Fiesta \citep{cunningham2013} and EFIT\texttt{++} \citep{lao1985, berkery2021, kogan2022} on MAST-U plasma discharges \citep{pentland2024} and has been used to emulate plasma scenarios for plasma control \citep{agnello2024}.
Most importantly, and like almost all other equilibrium codes, FreeGSNKE currently returns a single solution to the GS equation upon simulation.

To systematically search for multiple solutions to the GS problem, we will use the deflated continuation algorithm proposed by \cite{farrell2016}.
This algorithm is able to identify multiple solutions to a PDE (when varying a parameter) by modifying the nonlinear problem to guarantee non-convergence to known solutions, under certain conditions. This means that when the nonlinear solver (e.g.~Newton's method) is applied again, if the solver converges then it has discovered another, distinct solution.
This algorithm enables the user to construct (possibly disconnected) bifurcation diagrams, which show how the number of solutions change with the PDE parameter being varied.
It has already proven successful in a number of different application areas, identifying hundreds of new stable/unstable equilibria for magnetic rotors \citep{cisternas2024} and discovering multiple experimentally-observed solutions in smectic liquid crystals \citep{xia2021}.
We will explore the static forward GS problem, FreeGSNKE, and deflated continuation in more detail below.

By harnessing the capabilities of both FreeGSNKE and deflated continuation, we will search for multiple solutions by varying certain parameters in the (toroidal) plasma current density function of the GS equation. 
We also assess whether varying the current in one of the active poloidal field coils (that control the shape of the plasma) also affects the number of solutions found. 
We will do this for a single time slice during the flat-top (steady state) phase of a MAST-U shot and generate bifurcation diagrams for these varying parameters.

\subsection{Related work} \label{sec:related_work}

Early studies of the GS equation primarily focused on proving the existence and uniqueness of solutions in simplified domains, under restricted boundary conditions, and with reduced plasma current density profiles.
Under these restrictions, one can formulate an eigenvalue problem from the GS equation with a free boundary\footnote{These problems typically assign a free, but constant, value to the flux on the computational boundary, whereas in real-world GS problems, the flux is allowed to vary (spatially) on the boundary, see \eqref{eq:Grad--Shafranov_BC}.} to prove solution existence, uniqueness (for cases with small eigenvalues), and non-uniqueness (for cases with larger eigenvalues).
This was work initially carried out by \cite{temam1975, temam1977} and \cite{schaeffer1977}, then, in a slightly more general setting, by \cite{ambrosetti1980} and \cite{berestycki1980}.
Similar problems were later revisited by \cite{bartolucci2021} and \cite{jeyakumar2021}.

In slightly more real-world settings, some work has been done to provide analytical derivations of multiple GS equilibria (again, under simplifying conditions and with constant unknown flux on the boundaries). 
For example, in a cylindrical plasma with polynomial current profiles (similar to those used in modern equilibrium codes), \cite{turnbull1984} identified settings in which variations in certain parameters (e.g.\ plasma current, flux on the conducting wall, wall radius) yield up to two GS solutions.

In a similar setting, but with stepped current density profiles, \cite{ilgisonis2004} demonstrate that variations in the magnitude and location of the stepped profile leads to a fold bifurcation, resulting in up to three solutions.
Furthermore, they suggest that bifurcations in GS equilibria may not appear in numerical simulations due to the current density normalisation process (present in almost all equilibrium solvers for numerical stability), which they say restricts the poloidal flux values that solvers can identify.
In \cref{sec:GS_problem}, we discuss this process and provide an experiment in \cref{sec:numerics} that demonstrate this statement might hold for Picard-based solvers but not for Newton-based solvers.

Other works include that of \cite{solano2004} in which the criticality (bifurcation) of GS solutions is discussed for polynomial current density profiles. 
Similarly to \cite{ilgisonis2004}, \cite{schnack2009} also finds evidence of up to two GS equilibria in the case of a tall thin plasma column with stepped current density profiles. 

Despite these theoretical studies and the widespread use of various equilibrium codes, there has been surprisingly little numerical investigation into the existence of multiple solutions to the GS equation for free--boundary equilibria in settings relevant to real-world tokamak operations.
A step towards this goal was made by \cite{ham2024}, in which they identify multiple solutions to a fixed--boundary GS problem numerically.
Using the Firedrake finite element package \citep{firedrake2023}, they are able to use deflated continuation to consider much more physically realistic plasma boundary shapes and systematically search for multiple solutions by varying parameters such as the aspect ratio, elongation, and triangularity.
Despite this progress, we still lack an investigation for truly free--boundary GS formulations that incorporate the integral boundary conditions, more realistic internal plasma current profiles, and external conductor currents used in real-world tokamak experiments.

\subsection{Paper structure}
The rest of this paper is organised as follows. 
In \cref{sec:GS_problem} we present the static forward GS problem, how it is solved in FreeGSNKE, and which parameters we will vary during the search for multiple solutions. 
In \cref{sec:defcon} we outline the deflated continuation algorithm, remarking on a number of algorithmic parameter choices that need to be made in order to efficiently search for solutions.
The multiple solutions found using FreeGSNKE and deflated continuation will be presented and analysed in \cref{sec:numerics}. 
Finally in \cref{sec:discussion} we discuss what these results mean for the future of GS equilibrium simulation and propose some ideas for future work in this area.

\section{The static forward free--boundary Grad--Shafranov problem} \label{sec:GS_problem}

The GS equation is a nonlinear elliptic PDE used ubiquitously in plasma equilibrium modelling for describing the (static, time-independent) balance between magnetic and plasma pressure forces in ideal MHD equilibria \citep{grad1958,shafranov1958}. 
It governs the poloidal flux $\psi(R,Z)$, which has units $[\text{Weber}/ 2\pi]$, within a two-dimensional cross-section of a toroidally ($\phi$) symmetric tokamak device and is given by 
\begin{align} \label{eq:Grad--Shafranov}
    \Delta^* \psi = -\mu_0 R J_{\phi}(\psi,R,Z), \quad (R,Z) \in \Omega,
\end{align}
where $\Delta^* \defeq R \partial_R R^{-1} \partial_R + \partial_{ZZ}$ is a linear elliptic operator and $\mu_0 = 4\pi \times 10^{-7} \ [N/A^2]$ is the magnetic permeability of free space. 
Here, $(R,\phi,Z)$ denotes the cylindrical coordinate system. 
\begin{figure}[t!]
    \centering
    \includegraphics[width=0.8\linewidth]{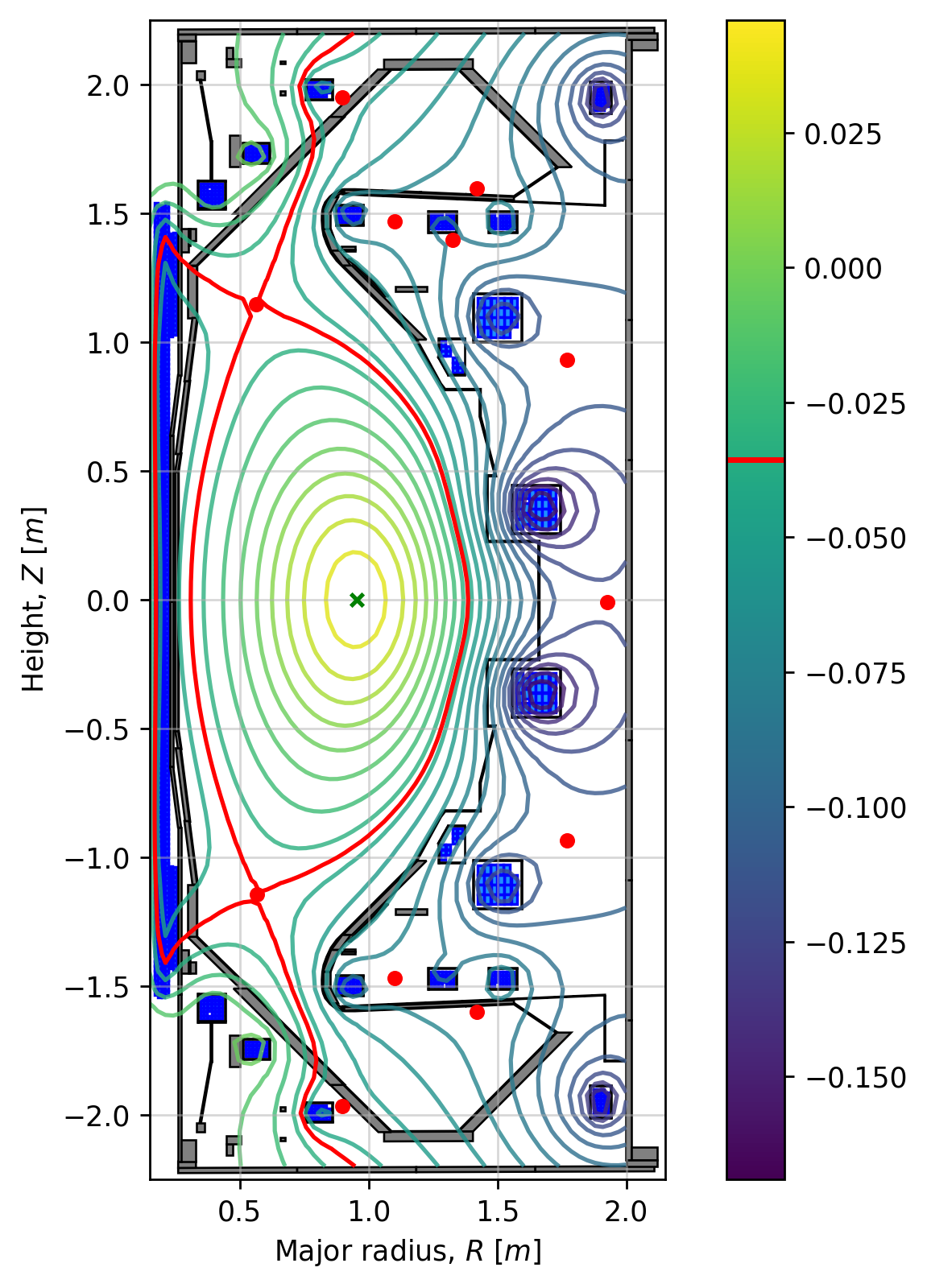}
    \caption{FreeGSNKE-simulated equilibrium of MAST-U shot 45272 ($t = 0.79854$s) with $\psi$ contours (see colour bar) shown in domain $\Omega = [0.06, 2] \times [-2.2, 2.2]$.
    Key features include the plasma region $\Omega_p$, which is enclosed by the last closed flux surface (solid red), the X-points (red dots), and the magnetic axis (green cross).
    Also shown are the twelve active poloidal field coils (dark blue), the passive structures (dark grey), and the wall/limiter (solid black).}
    \label{fig:MAST-U}
\end{figure}

The poloidal flux $\psi$ is the sum of two terms $\psi \defeq \psi_p + \psi_c$, where $\psi_p$ and $\psi_c$ are contributions from the plasma and the (toroidally symmetric) conducting metal structures external to the plasma\footnote{The external conductors are the active poloidal field coils (whose currents are used to shape and control the plasma) and the passive conducting structures (whose currents are induced by the plasma and the active coils).}, respectively.
The toroidal current density $J_{\phi}(\psi, R, Z) \defeq J_{p}(\psi, R, Z) + J_{c}(R, Z)$ also contains contributions from both the plasma $J_{p}$ and the external conductors $J_{c}$.
The dependence on $\psi$ is where part of the nonlinearity in the static forward GS problem originates.
The computational domain is a pre-specified rectangular grid denoted by $\Omega \defeq \Omega_{p} \cup \Omega_{p}'$, where $\Omega_p$ defines the plasma region (whose boundary is to be determined) and $\Omega_p'$ refers to its complement (see \cref{fig:MAST-U}).

An integral (Dirichlet) free-boundary condition accompanies \eqref{eq:Grad--Shafranov}:
\begin{align} \label{eq:Grad--Shafranov_BC}
    \psi \bigg\rvert_{\partial \Omega} = \int_{\Omega} G(R,Z;R',Z') J_{\phi}(\psi, R',Z') \ \mathrm{d}R' \mathrm{d}Z',
\end{align}
which gives the flux on the computational boundary $\partial \Omega$ produced by all non-zero toroidal current sources in $\Omega_p$ and $\Omega_c$ (see \eqref{eq:ext_currents} for external conductor currents).
The function $G$ is the (analytically) known Green's function for the operator $\Delta^*$ and links points on the boundary with the toroidal current sources. 
Refer to \citet[Sec.~3.1]{takeda1991} for further details on $G$ and to \citet[Chap.~4.6.4]{jardin2010} for how this integral can be calculated efficiently. 

It should be noted that the plasma boundary $\partial \Omega_p$ is defined by the (last) closed $(R,Z)$ contour of $\psi$ that passes through the X-point closest to the magnetic axis of the plasma\footnote{The X-point and magnetic axis (sometimes referred to as an O-point) are identified by finding the critical points of $\psi$ (see \citet[Sec.~5]{jeon2015}).}.
The (nonlinear) identification of $\partial \Omega_p$ is required in order to calculate $J_p$ (see \eqref{eq:plasma_current_density}) and therefore solve the forward GS problem \eqref{eq:Grad--Shafranov}--\eqref{eq:Grad--Shafranov_BC}.


\subsection{The toroidal current density}
Here, we outline how the plasma and external conductor current density contributions to $J_{\phi}$ are defined and, more crucially, highlight which of their parameters we vary later on in deflated continuation to identify multiple GS equilibria.

\subsubsection{Plasma current density}

The plasma current density is governed by the distribution of charged particles within the plasma and generates a poloidal magnetic field that contributes to the confinement of the plasma. 
It is non-zero only within $\Omega_{p}$, taking the form 
\begin{align} \label{eq:plasma_current_density}
     J_{p}(\psi, R, Z) = R \frac{\mathrm{d} p}{\mathrm{d} \psi} + \frac{1}{\mu_0 R} F \frac{\mathrm{d} F}{\mathrm{d} \psi}, \quad (R,Z) \in \Omega_{p},
\end{align}
where $p \defeq p(\psi)$ is the plasma pressure profile and $F \defeq F(\psi) = R B_{\phi}$ is the toroidal magnetic field profile ($B_{\phi}$ is the toroidal component of the magnetic field).

The full specification of the forward GS problem \eqref{eq:Grad--Shafranov}--\eqref{eq:Grad--Shafranov_BC} requires specifying $p'$ and $FF'$ to determine the internal profiles of the pressure and toroidal current within the plasma\footnote{Fitting these profiles is part of solving the GS equilibrium \emph{reconstruction} problem---see \citet[Sec.~4.7]{jardin2010} for an introduction.}.
For real-world tokamak plasmas, these can take various nonlinear forms, making analytical progress difficult and numerical computation challenging.
\begin{figure}[t!]
    \centering
    \includegraphics[width=0.99\linewidth]{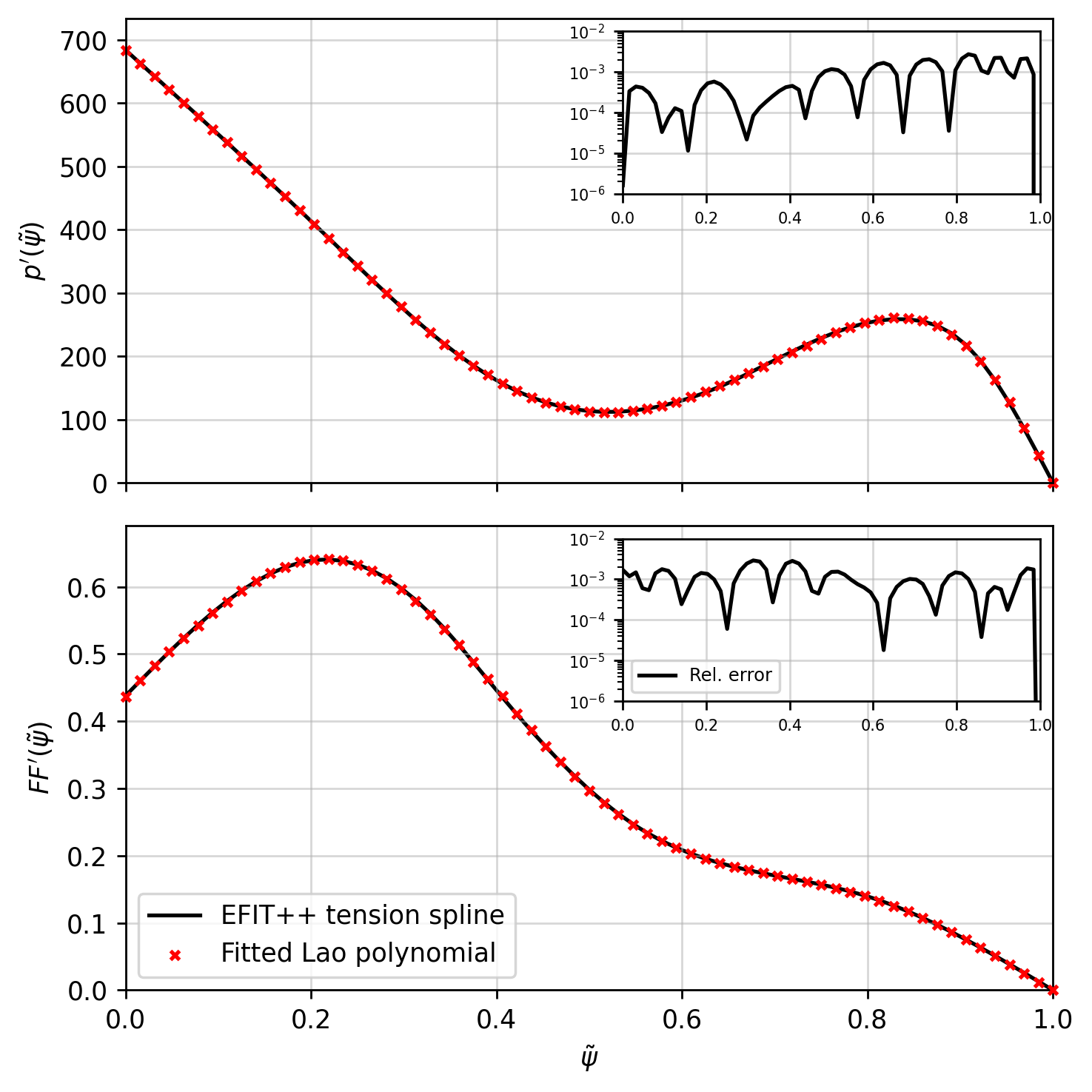}
    \caption{Plasma current density profiles $p'$ (top) and $FF'$ (bottom) used to simulate the equilibrium in \cref{fig:MAST-U}.
    Shown are the original tension spline profiles obtained from the EFIT\texttt{++} reconstruction (solid black) and the profiles fit using the Lao polynomial parameterisation \eqref{eq:lao_profiles} (red dots). 
    The inset plots display the relative error between the two different parameterisations.}
    \label{fig:profiles}
\end{figure}

Here, we use the (Lao) polynomial profile parametrisation:
\begin{equation} \label{eq:lao_profiles}
\begin{aligned}
    \frac{\mathrm{d} p}{\mathrm{d} \tilde{\psi}} &= \sum_{i=0}^{n_p} \alpha_i \tilde{\psi}^i - \bar{\alpha} \tilde{\psi}^{n_p + 1} \sum_{i=0}^{n_p} \alpha_i, \\
    F \frac{\mathrm{d} F}{\mathrm{d} \tilde{\psi}} &= \sum_{i=0}^{n_F} \beta_i \tilde{\psi}^i - \bar{\beta} \tilde{\psi}^{n_F + 1} \sum_{i=0}^{n_F} \beta_i.
\end{aligned}
\end{equation}
where $\alpha_i, \beta_i \in \Reals$ are coefficients, $\bar{\alpha}, \bar{\beta} \in \{0, 1\}$ are Booleans (governing edge conditions at $\tilde{\psi}=1$), and $n_p, n_F \in \Natural \cup {0}$ dictate the polynomial orders \citep{lao1985}.
Also note
\begin{align} \label{eq:normalised_psi}
    \tilde{\psi} = \frac{\psi - \psi_a}{\psi_b - \psi_a} \in [0,1],
\end{align}
defines the normalised poloidal flux ($\psi_a$ and $\psi_b$ being the flux on the magnetic axis and plasma boundary, respectively).

In the numerical experiments with deflated continuation (to follow in \cref{sec:numerics}) we will vary the coefficients $\alpha_i, \beta_i$.
\cref{fig:profiles} displays the profiles used to obtain the equilibrium in \cref{fig:MAST-U}.
In addition, we will vary the total plasma current $I_p$, whose value is not strictly required as an input to solve \eqref{eq:Grad--Shafranov}--\eqref{eq:Grad--Shafranov_BC}, but is often prescribed to normalise \eqref{eq:plasma_current_density}. 
This simply involves multiplying the right hand side of \eqref{eq:plasma_current_density} by 
\begin{align} \label{eq:normalisation}
    I_p \left( \int_{\Omega_p} J_p(\psi, R, Z) \ \mathrm{d}R \mathrm{d}Z \right)^{-1}.
\end{align}

\subsubsection{External conductor current density}
To close the system we must also specify the current density $J_c$ within the metal conductors external to the plasma in the tokamak, i.e.\ the active poloidal field coils and the passive (non-active) structures, as shown in \cref{fig:MAST-U}.
These conductors generate magnetic fields for shaping and controlling the plasma position and stability.
The current density produced by $N_{c}$ external conductors is modelled as
\begin{align} \label{eq:ext_currents}
     J_{c}(R, Z) &= \sum_{j=1}^{N_{c}} \frac{I_j^{c}(R,Z)}{A_j^c}, \quad (R,Z) \in \Omega, \\
     I_j^c(R,Z) &= \begin{cases}
    I_j^c & \ \text{if} \ (R,Z) \in \Omega_j^c, \vspace*{0.00cm} \\ 
    0 & \ \text{elsewhere}, \vspace*{0.00cm} \nonumber
\end{cases}
\end{align}
where $\Omega_j^c$, $I_j^{c}$, and $A_j^c$ are the domain region, current, and cross-sectional area of the $j$\textsuperscript{th} conductor, respectively.
Note that this term can be calculated explicitly before solving \eqref{eq:Grad--Shafranov}--\eqref{eq:Grad--Shafranov_BC}.

In \cref{sec:numerics}, we will investigate how the equilibrium solutions vary as we change a current in one of the active poloidal field coils in MAST-U.

\subsection{Numerical solution} \label{sec:numerical_solution}

Solving \eqref{eq:Grad--Shafranov}--\eqref{eq:Grad--Shafranov_BC} with FreeGSNKE requires a number of different inputs specific to the MAST-U machine.
Firstly, we require a machine description of MAST-U that includes the position, size, orientation, and polarity of the active poloidal field coils, passive structures, and the limiter/wall (that will confine the boundary of the plasma during the simulation).
Then, to simulate a specific equilibrium (at a given time slice of a shot) we require the plasma profile coefficients/parameters for \eqref{eq:lao_profiles}, the total plasma current $I_p$ for the normalisation in \eqref{eq:normalisation}, and the currents measured in the external conductors for \eqref{eq:ext_currents}.
The input data required to do this comes from an EFIT\texttt{++} reconstruction of the plasma equilibrium.

In this paper, we use data obtained from a magnetics plus motional Stark effect EFIT\texttt{++} reconstruction.
This reconstruction code uses measured coil currents, plasma current, magnetic fields, and motional Stark effect data in order to find the ``best'' fit for the aforementioned parameters \citep{conway2010}.
We should note that this type of EFIT\texttt{++} reconstruction actually fits parameters to the tension spline parameterisation of the $p'$ and $FF'$ profiles (see \citet[App.~A]{pentland2024}), however, these parameters are difficult to use within the deflated continuation framework.
To suit our needs, we instead fit the $\alpha$ and $\beta$ coefficients of the Lao polynomial profiles \eqref{eq:lao_profiles} to the tension spline profiles from EFIT\texttt{++}---this fit (and the relative errors) can be seen in \cref{fig:profiles}.
For profiles of this complexity, we required polynomials up to order $n_p=n_F=9$. 

In FreeGSNKE, the static forward GS problem \eqref{eq:Grad--Shafranov}--\eqref{eq:Grad--Shafranov_BC} is solved in its residual form
\begin{align} \label{eq:GS_residual}
    F(\psi; \lambda) \equiv \Delta^* \psi + \mu_0 R J_{\phi} (\psi, R, Z; \lambda) = 0,
\end{align}
where we have slightly abused notation by omitting the boundary condition (though it is indeed applied).
Here $\lambda$ denotes the (scalar) parameter in $J_{\phi}$ that we will vary when searching for multiple solutions with deflated continuation.

Once discretised (using fourth-order finite differences), \eqref{eq:GS_residual} is solved using a Jacobian-free Newton-Krylov (NK) method (see \citet[App.~1]{amorisco2024}) with an appropriate initial guess for the plasma flux $\psi_p$ (recall $\psi_c$ is known a priori to simulation).
If no initial guess for $\psi_p$ is provided, FreeGSNKE generates one by default with ellipse-shaped flux contours, the magnitude of which are adaptively scaled up such that the total flux produces a magnetic axis and an X-point within the confining limiter geometry. 
The NK method then iterates until a relative convergence tolerance
\begin{align} \label{eq:tolerance}
    \frac{\max | F(\psi; \lambda) |}{\max (\psi) - \min (\psi)} < \varepsilon,
\end{align}
is met, returning a \emph{single} solution $\psi = \psi_p + \psi_c$ to the problem. 

Searching for multiple solutions to \eqref{eq:GS_residual} by using a set of different initial guesses $\psi_p$ is difficult for a number of reasons.
Firstly, it is difficult to efficiently generate such a set and, even if we could, there is no way to guarantee that the input space would be well covered by a given number of them.
Secondly, many of these initial guesses would most likely converge to the same solution and many may not converge at all.
This task is both cumbersome and computationally inefficient, hence we now explain how we can systematically search for multiple \emph{distinct} solutions while varying the parameter $\lambda$ in \eqref{eq:GS_residual} using the deflated continuation algorithm.

\section{Deflated continuation} \label{sec:defcon}

To search for multiple solutions to the static forward GS problem, as well as potential bifurcation points in parameter space, we use the deflated continuation algorithm first proposed by \cite{farrell2016}.

\subsection{How it works}
For simplicity we consider the discretised problem. The purpose of deflated continuation is to locate solutions $u \in \Reals^m$ to
\begin{align} \label{eq:residual}
    F(u; \lambda) = 0,
\end{align}
where $F \colon \Reals^m \times \Reals \rightarrow \Reals^n$ is a nonlinear function that represents the residual of a PDE problem (e.g.\ \eqref{eq:GS_residual}) and $\lambda \in \Reals$ is a parameter in the equations.
More specifically, it will locate a set of distinct solutions $\{ u_1^*, u_2^*, \ldots \}$ to \eqref{eq:residual} for \emph{each} of the parameter values considered, typically $L+1$ equally spaced ($\Delta \lambda$) values of $\lambda$ in a chosen interval $[\lambda_0, \lambda_L]$.
An important advantage over other methods for computing multiple solutions (e.g.\ pseudo-arclength continuation and branch switching) is that deflated continuation is able to compute disconnected bifurcation diagrams in which solutions on different branches may not meet at bifurcation points (see \citet[Fig.~1.2]{farrell2016} for an example).
It can do this by combining the power of both \emph{deflation} and \emph{continuation}. 

Suppose we have found a solution $u_1^*$ to \eqref{eq:residual}, for fixed $\lambda$, using a suitable nonlinear root finding method (e.g.~NK).
To try to identify more solutions, we can use the deflation technique \citep{brown1971,farrell2015} to modify the operator \eqref{eq:residual} such that we instead solve
\begin{align} \label{eq:def_residual}
    M(u; u^*_1) F(u; \lambda) = 0,
\end{align}
where 
\begin{align} \label{eq:def_operator}
    M(u; u^*_1) = \left( \frac{1}{\| u - u^*_1 \|_2^{p}} \right) + \sigma,
\end{align}
is the \emph{deflation operator} with power and shift parameters $p > 0$ and $\sigma > 0$, respectively ($\| \cdot \|_2$ denotes the Euclidean norm).
Under mild regularity conditions, solving \eqref{eq:def_residual} ensures the nonlinear solver does not return the known solution $u^*_1$ (using the same initial guess as before) but rather a distinct solution $u_2^*$ (if the method converges, which is not guaranteed).
The problem in \eqref{eq:def_residual} can subsequently be deflated again using known solutions $\{ u^*_1, \ldots, u_N^* \}$ such that
\begin{align} \label{eq:def_residuals}
    \prod_{i=1}^{N} M(u; u^*_i) F(u; \lambda) = 0,
\end{align}
is solved until no more are found within a specified number of nonlinear iterations.

To initialise the algorithm at the first $\lambda$, \eqref{eq:residual} is solved using the chosen nonlinear solver and the solution(s)\footnote{If any other solutions are known or found, they can also be used at this stage.} recorded. 
Then, for $\lambda + \Delta \lambda$, deflated continuation carries out two separate stages for solving \eqref{eq:residual}: continuation and exploration.
\begin{description}
    \item[Continuation stage:] Each known solution (from step $\lambda$) is used as an initial guess in the nonlinear solver to try to find the corresponding solution at $\lambda + \Delta \lambda$.
    For each solution successfully continued\footnote{The implicit function theorem states that a solution branch should only cease to exist if the Fr{\'e}chet derivative of the residual function (at that state) is zero. If not, then it suggests that the nonlinear solver has failed due to lack of iterations, instability, etc.}, the new solution is used to deflate the residual function as in \eqref{eq:def_residuals}.
    \item[Exploration stage:] Again, each known solution (from step $\lambda$) is used as an initial guess to solve \eqref{eq:def_residuals} to try to locate additional new solutions (at $\lambda + \Delta \lambda$).
    Again, if any new solutions are found, they are used to deflate \eqref{eq:def_residuals} before considering the next initial guess.
\end{description}
All solutions found for $\lambda + \Delta \lambda$ are then stored, ready for use when considering the next value of $\lambda$.

\subsection{Numerical implementation}

The original implementation of deflated continuation, developed by \cite{defcon2016}, is built using the Firedrake \citep{firedrake2023} and FEniCS \citep{baratta2023} libraries.
Here, we use a purpose-built Python implementation, tailored specifically for use with FreeGSNKE.
This is necessary as FreeGSNKE uses its own finite difference scheme and its own purpose-built NK solver for tackling \eqref{eq:residual}.

Before each simulation, we need to choose a few key parameters within both deflated continuation and FreeGSNKE.
When solving \eqref{eq:residual} (or \eqref{eq:def_residuals}) with the NK solver, we use a relative convergence tolerance of $\varepsilon = 10^{-6}$ in \eqref{eq:tolerance}, a maximum of $150$ NK iterations, and a scaled Newton step size of $1.2$ (all other settings are FreeGSNKE defaults).
Given there is no systematic way of choosing the free parameters $(p,\sigma)$ in \eqref{eq:def_operator} (see \cite{farrell2015} for a discussion on this), we set them independently for each experiment in \cref{sec:numerics} depending on which combination works well---the best default choice was found to be $(p,\sigma) = (1, 0.05)$.
For each parameter $\lambda$ under consideration, the interval $[\lambda_0, \lambda_L]$ and the chosen step size $\Delta \lambda$ will vary.
In addition, we explore each side of the starting value $\lambda_0$, i.e. we explore $[\lambda_{-L}, \lambda_0]$ using step $-\Delta \lambda$. 

\section{Searching for multiple solutions} \label{sec:numerics}

In this section, we carry out our search for multiple solutions to the static forward GS problem laid out in \cref{sec:GS_problem} using deflated continuation.
Here, we solve \eqref{eq:GS_residual} at time $t=0.79854$s of MAST-U shot $45272$ using $m = 65^2 = 4425$ spatial grid points (recall the equilibrium in \cref{fig:MAST-U} and profiles used in \cref{fig:profiles}).
This is a double-null plasma with a flat-top current of approximately $750$kA (heated by two neutral beam injection systems) and a conventional divertor configuration. 

The inputs required to solve for the equilibrium at this time slice are described in \cref{sec:numerical_solution} and will be used as the starting parameters ($\lambda_0$) in our deflated continuation experiments. 
The parameters we can vary for this equilibrium are the: 
\begin{itemize}
    \item coefficients $\alpha_i$, $\beta_i$ in the $9^{\text{th}}$ order Lao polynomials \eqref{eq:lao_profiles};
    \item plasma current $I_p$ in \eqref{eq:normalisation};

    \item coil currents $I^c_j$, $j \in \{1,\ldots,12 \}$, in \eqref{eq:ext_currents}. 
\end{itemize}
We will investigate whether any bifurcations occur when varying some of these parameters (separately) over a suitable interval.
To initialise deflated continuation at $\lambda_0$, we use the plasma flux $\psi_p$ found by FreeGSNKE during a standard solve (i.e. what we find in \cref{fig:MAST-U}).

To plot bifurcation diagrams for vector solutions (e.g. $\psi$), it is common to calculate scalar-valued \emph{functionals} related to the solution.
While many functionals of $\psi$ are available (e.g.\ inner/outer midplane radii, strikepoints, X-points), we find that the most informative are $\psi_a$ and $\psi_b$ (recall \eqref{eq:normalised_psi}).

\begin{figure}[t!]
    \centering
    \begin{subfigure}{0.9\linewidth}
        \includegraphics[width=0.99\textwidth]{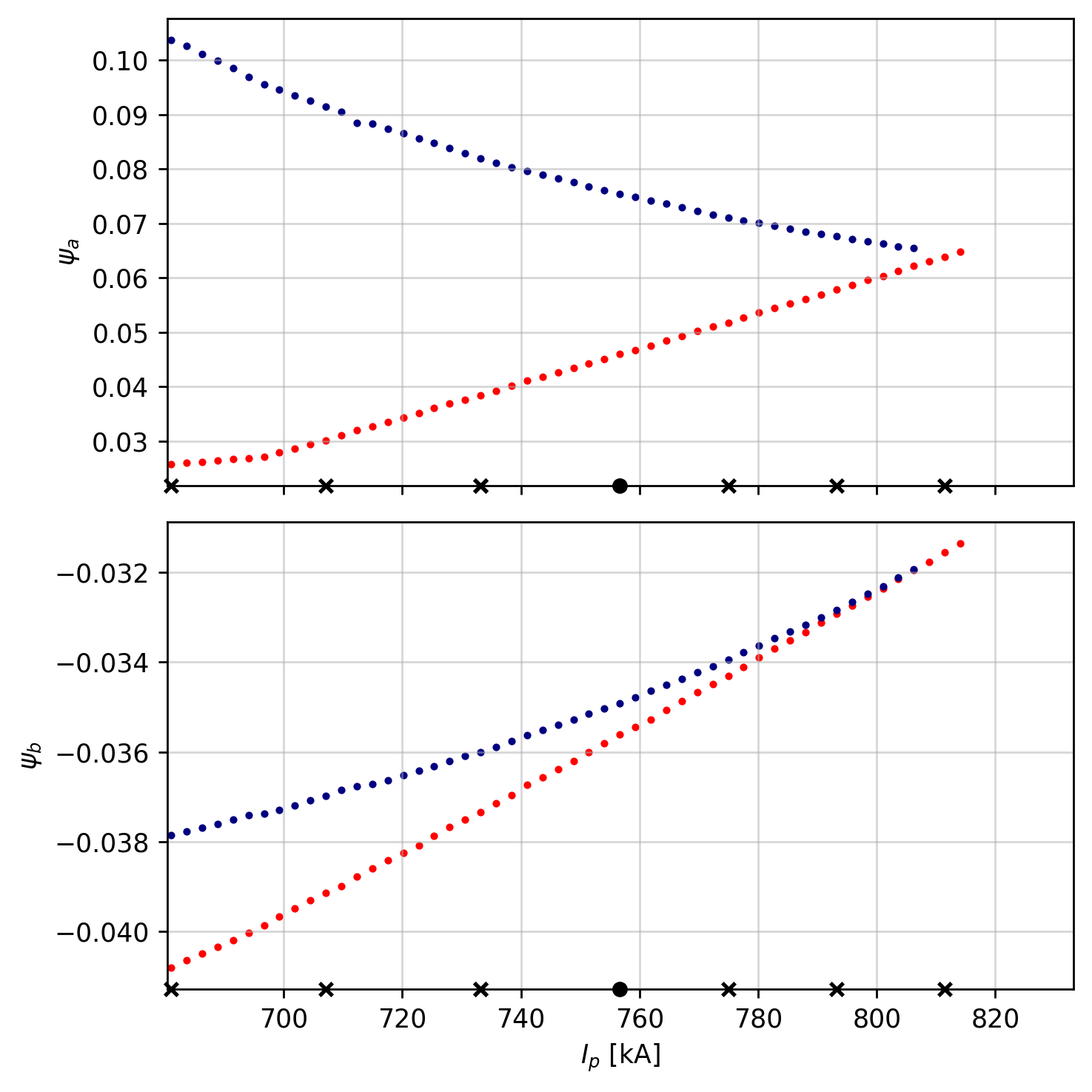}
    \end{subfigure}
    \caption{Bifurcation diagrams for $\psi_a$ (top) and $\psi_b$ (bottom) when varying plasma current $I_p$.
    Different solution branches are indicated by different colours and the initial $\lambda_0$ value is indicated by the black dot on the x-axis. 
    The solutions in \cref{fig:Ip_solutions} are plot in the same colours at the values of $I_p$ indicated by the black crosses (and the dot) on the x-axis.
    }
    \label{fig:Ip}
\end{figure}
\begin{figure*}[t!]
    \centering
    \begin{subfigure}{0.99\linewidth}
        \includegraphics[width=0.99\textwidth]{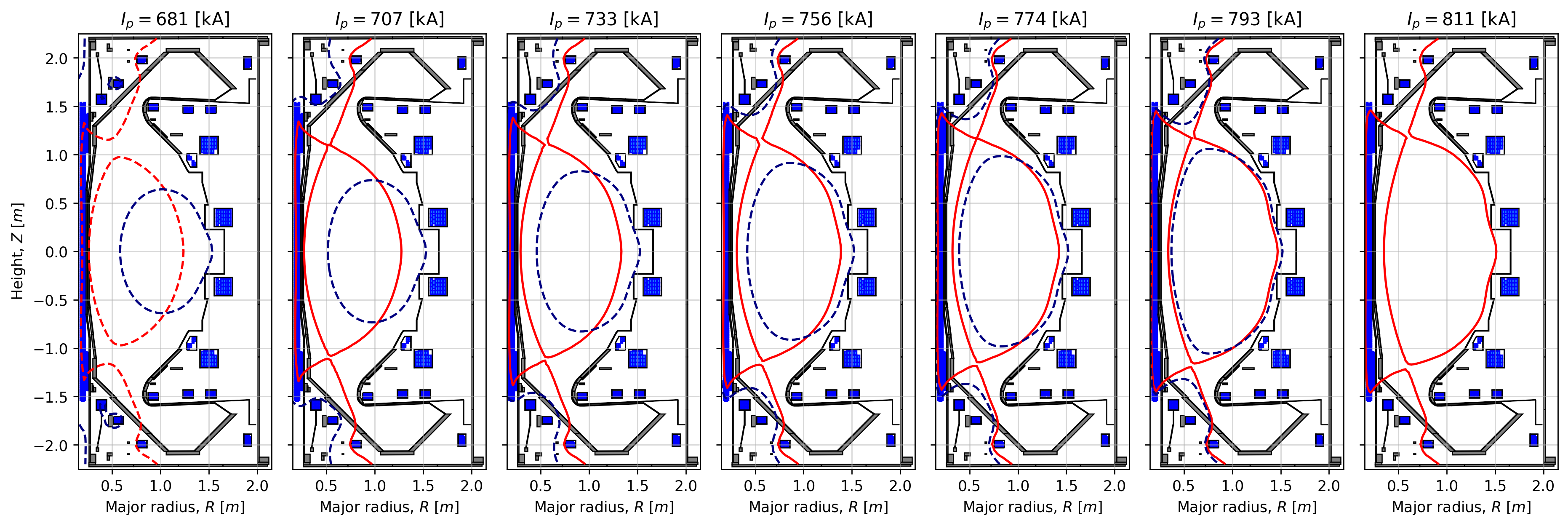}
    \end{subfigure}
    \caption{Separatrices of the multiple equilibria (red and blue) at increasing values of $I_p$ (whose values are indicated by the black crosses and dot on the x-axis of \cref{fig:Ip}).
    A dashed separatrix line indicates that the plasma is limited (i.e.~touching the wall) while a solid line indicates it is diverted.
    }
    \label{fig:Ip_solutions}
\end{figure*}
\begin{figure*}[t!]
    \centering
    \begin{subfigure}{0.42\linewidth}
        \includegraphics[width=0.99\textwidth]{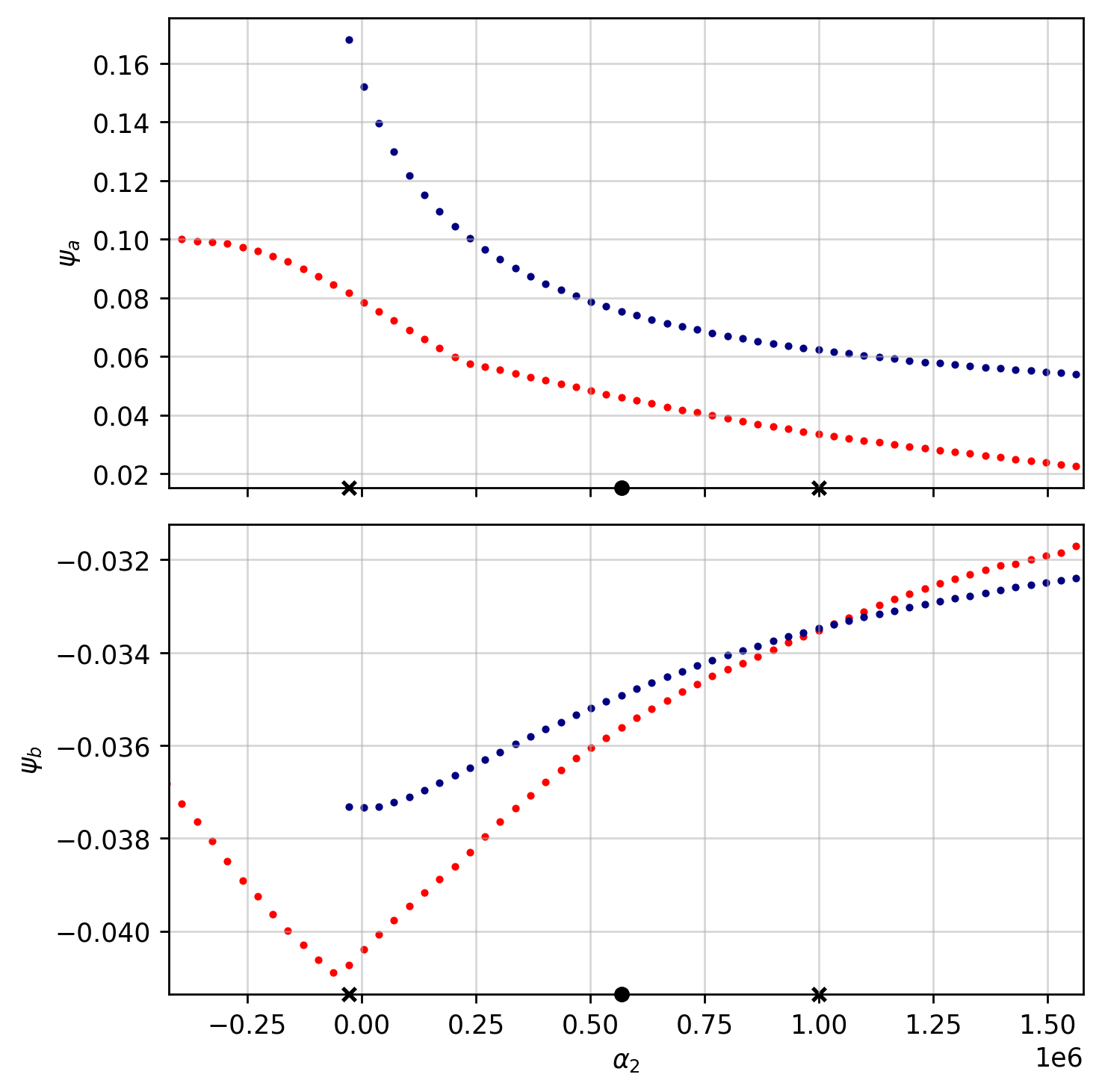}
    \end{subfigure}
    \begin{subfigure}{0.54\linewidth}
        \includegraphics[width=0.99\textwidth]{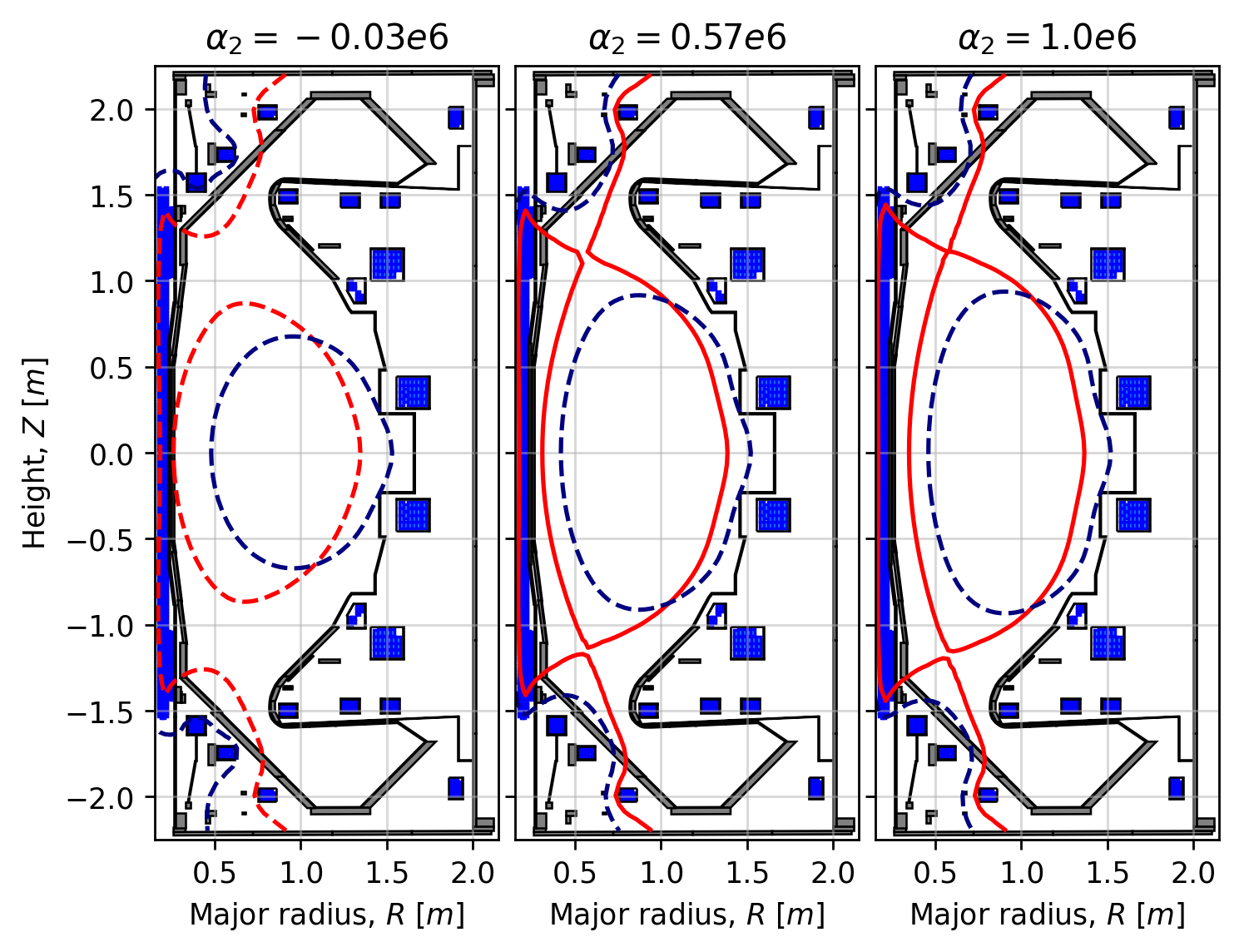}
    \end{subfigure}
    \caption{Left: bifurcation diagrams for $\psi_a$ and $\psi_b$ (left) when varying $\alpha_2$.
    Different solution branches are indicated by different colours and the initial $\lambda_0$ value is indicated by the black dot on the x-axis.
    Right: separatrices of the multiple equilibria (red and blue) at increasing values of $\alpha_2$ (whose values are indicated by the black crosses and dots on the x-axis of the left panel).
    A dashed separatrix line indicates that the plasma is limited (i.e.~touching the wall) while a solid line indicates it is diverted.
    }
    \label{fig:alpha}
\end{figure*}

\subsection{Experiments}

\subsubsection{Varying $I_p$}

The first parameter we test with deflated continuation is $\lambda = I_p$.
Starting with the initial value $\lambda_0 \approx 756$kA, we run deflated continuation both ``forward'' and ``backward'' to cover a range of $I_p$ values up to $10\%$ either side of $\lambda_0$.
In \cref{fig:Ip}, we can see that deflated continuation reveals two distinct solution branches that diverge from one another as $I_p$ decreases away from $\lambda_0$ and converge together above it.
The lower panel (displaying $\psi_b$), shows the branches merging around $I_p \approx 806$kA and the remaining solution branch being lost beyond $I_p \approx 814$kA.

This process can be seen more clearly when we plot the separatrices of the solutions $\psi$ at a number of different $I_p$ values in \cref{fig:Ip_solutions}.
For $\lambda_0$, the centre panel displays both a deeply confined diverted plasma equilibrium (i.e.~the one found initially by FreeGSNKE in \cref{fig:MAST-U}) and a more shallowly confined limited plasma on the outboard side.
Each separatrix corresponds to the respective solution branch colour shown in \cref{fig:Ip}.
As $I_p$ increases from left to right in \cref{fig:Ip_solutions}, we see the size of each plasma core increasing dramatically until the outboard limited solution merges into the diverted solution causing a bifurcation.
Beyond this point, the diverted equilibrium exists only for a small increase in $I_p$, expanding further until it can no longer be contained by the limiter/wall and is lost.
For $I_p \lessapprox 700$kA, the deeply confined equilibrium switches from a diverted plasma to a limited one.

We should note here that we also ran the same experiment (not shown) without the plasma current normalisation process (recall \eqref{eq:normalisation}) and found identical results to those shown here. 
This demonstrates that multiple solutions can indeed persist with or without the current normalisation process when using a Newton-based solver. 


\subsubsection{Varying $\alpha_i$ or $\beta_i$}

In \cref{fig:alpha}, we vary the coefficient $\alpha_2$ in the $p'$ profile \eqref{eq:lao_profiles} by up to $175\%$---notice the large magnitude of $\alpha_2$ in the left panels.
As before, we can see the two co-existing solutions we saw when varying $I_p$ (see right panel) and we observe that the outboard limiter solution is lost for $\alpha_2 \lessapprox -0.03\posE6$.
We can also see the upward curve on the lower branch (of the $\psi_a$ panel) for $\alpha_2 \lessapprox 0.25\posE6$ indicating the transition of the diverted solution into an inboard limited plasma.
In addition, we see that the two solutions co-exist with almost exactly the same $\psi_b$ value at $\alpha_2 \approx 1\posE6$ (in the $\psi_b$ panel).
Note that we do not see any bifurcations in this case.

\begin{figure*}[t!]
    \centering
    \begin{subfigure}{0.42\linewidth}
        \includegraphics[width=0.99\textwidth]{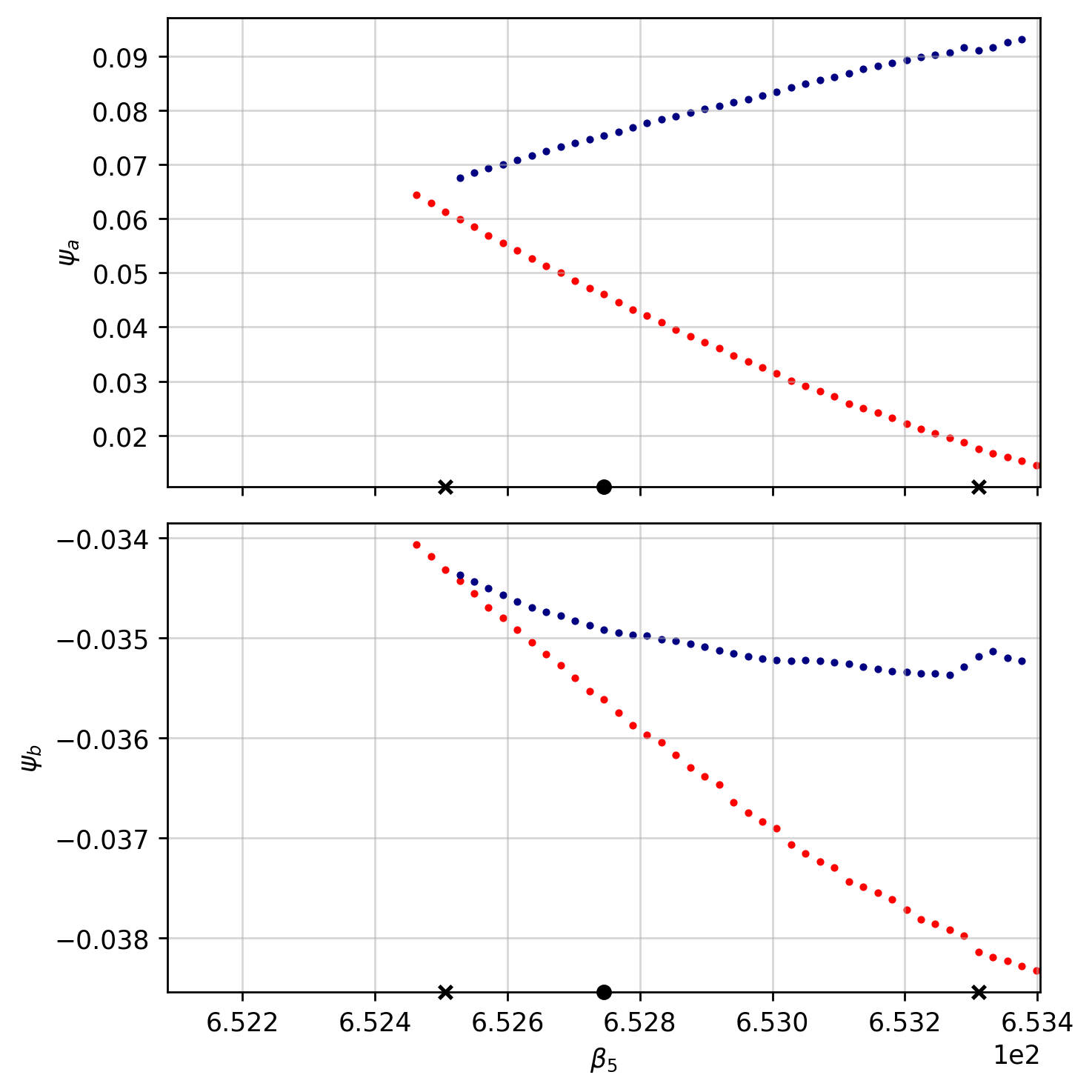}
    \end{subfigure}
    \begin{subfigure}{0.54\linewidth}
        \includegraphics[width=0.99\textwidth]{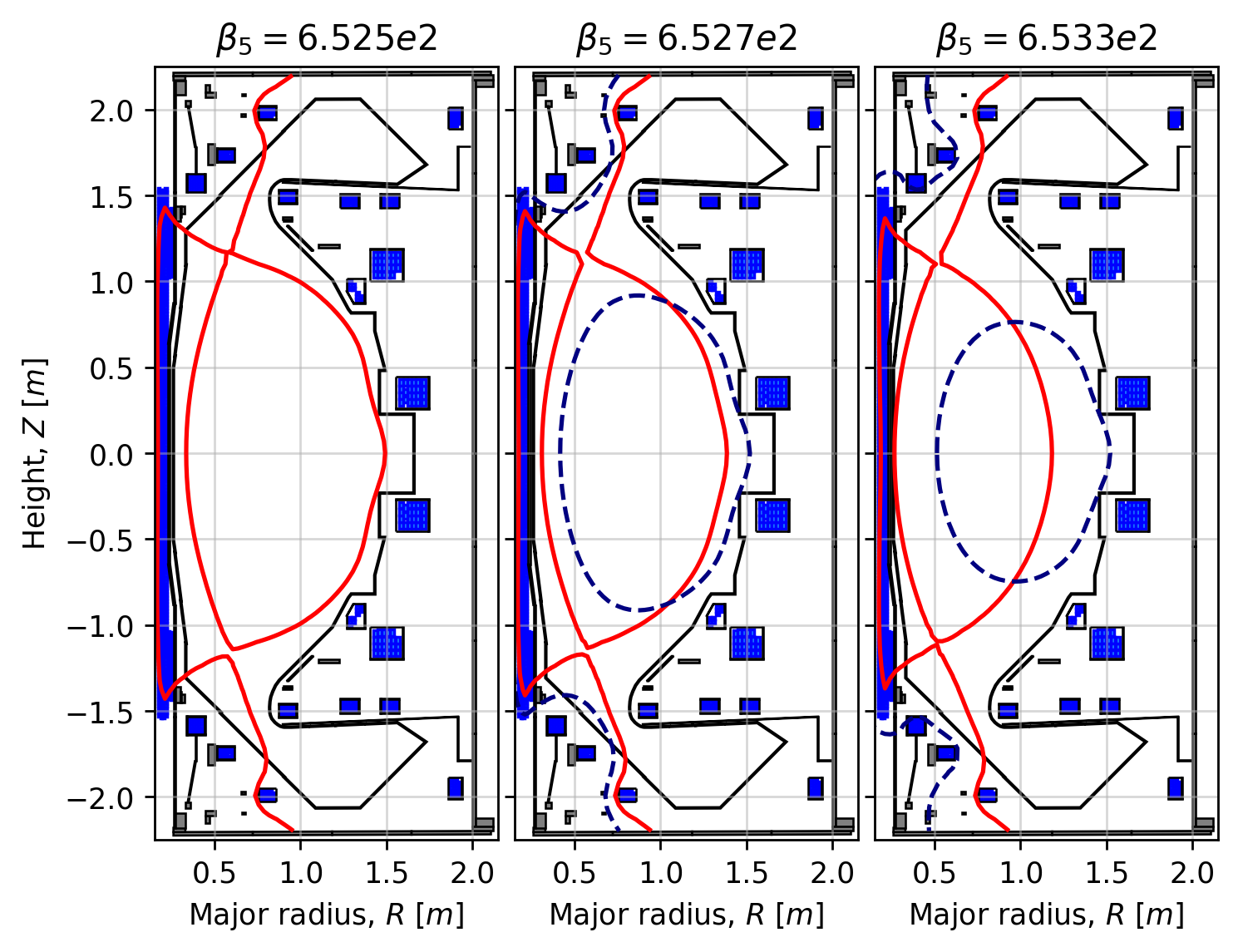}
    \end{subfigure}
    \caption{Left: bifurcation diagrams for $\psi_a$ and $\psi_b$ (left) when varying $\beta_5$.
    Different solution branches are indicated by different colours and the initial $\lambda_0$ value is indicated by the black dot on the x-axis.
    Right: separatrices of the multiple equilibria (red and blue) at increasing values of $\beta_5$ (whose values are indicated by the black crosses and dot on the x-axis of the left panel).
    A dashed separatrix line indicates that the plasma is limited (i.e.~touching the wall) while a solid line indicates it is diverted.
    }
    \label{fig:beta}
\end{figure*}

In \cref{fig:beta}, we vary the $\beta_5$ parameter and, again, see the two same solutions coexisting before bifurcating in a manner very similar to what we saw in with $I_p$ in \cref{fig:Ip} (though the solutions merge and are then lost as $\beta_5$ is decreases).

\subsubsection{Varying a coil current $I^c_j$}

\begin{figure*}[h!]
    \centering
    \begin{subfigure}{0.42\linewidth}
        \includegraphics[width=0.99\textwidth]{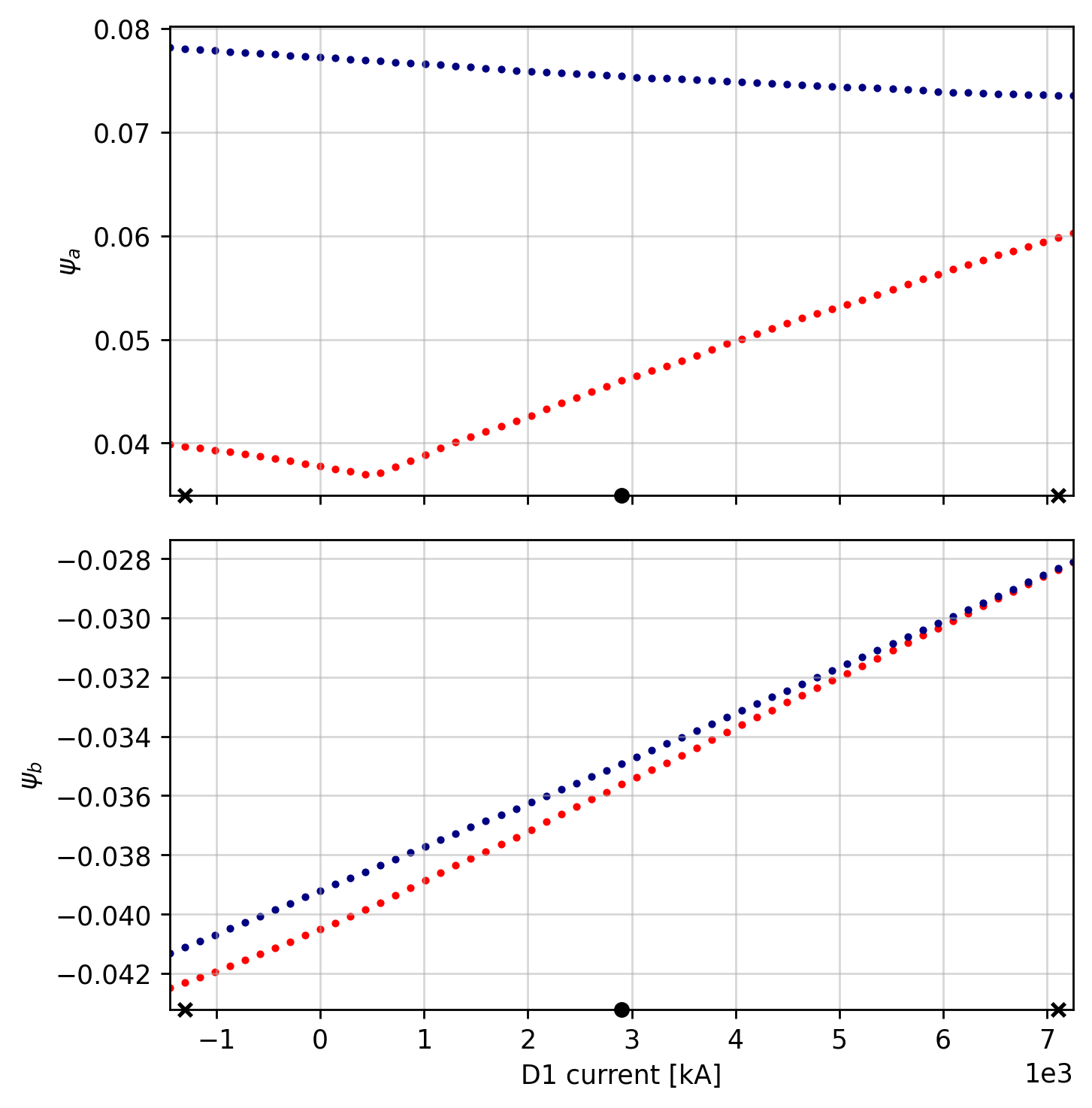}
    \end{subfigure}
    \begin{subfigure}{0.54\linewidth}
        \includegraphics[width=0.99\textwidth]{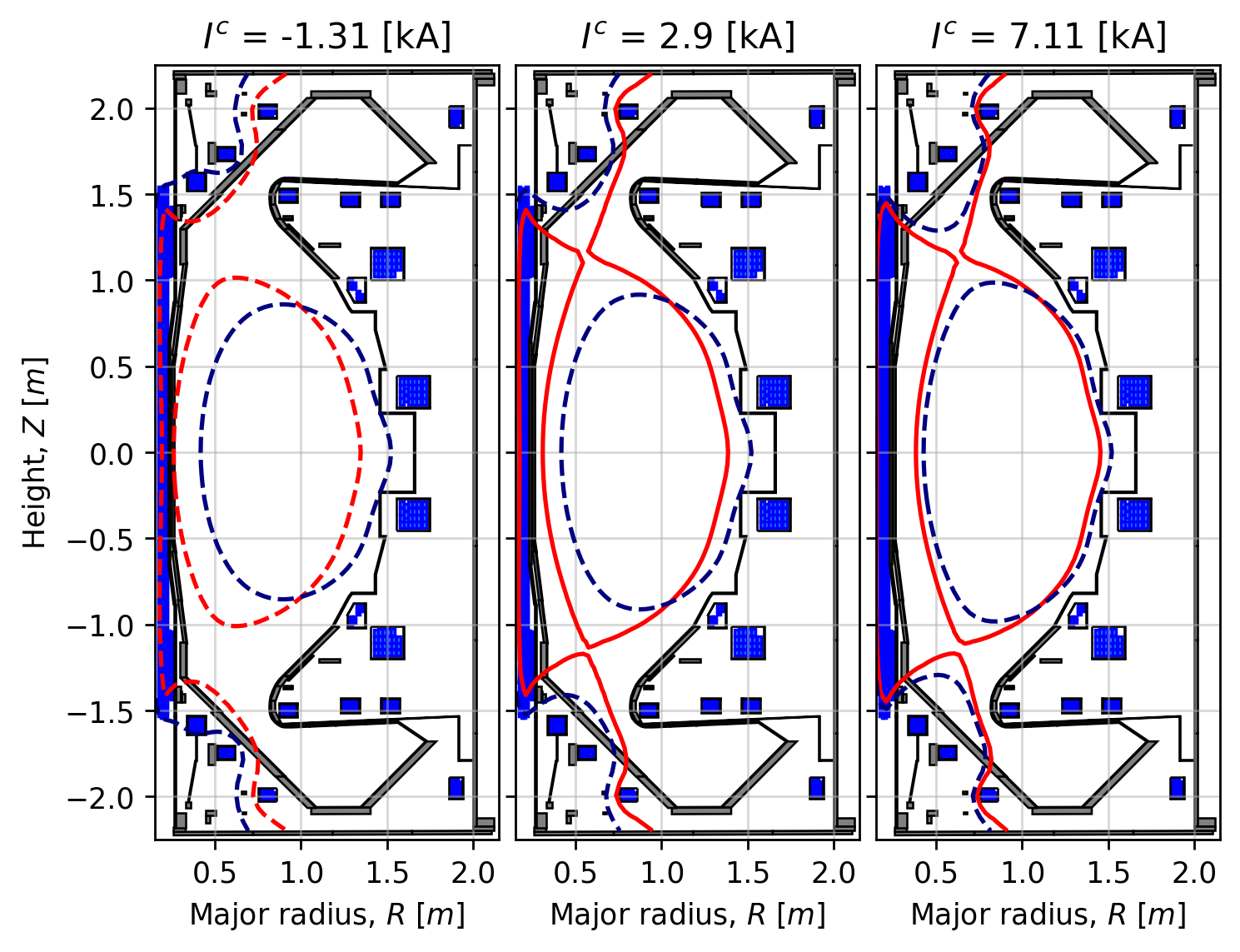}
    \end{subfigure}
    \caption{Left: bifurcation diagrams for $\psi_a$ and $\psi_b$ (left) when varying current in the D1 coil.
    Different solution branches are indicated by different colours and the initial $\lambda_0$ value is indicated by the black dot on the x-axis.
    Right: separatrices of the multiple equilibria (red and blue) at increasing values of D1 current (whose values are indicated by the black crosses and dot on the $x$-axis of the left panel).
    A dashed separatrix line indicates that the plasma is limited (i.e. touching the wall) while a solid line indicates it is non-limited.
    }
    \label{fig:current}
\end{figure*}

In this experiment, we run a deflated continuation simulation in which we instead vary the current in one of the poloidal field coils, specifically the D1 coil---see \citet[Fig.~1]{pentland2024}.
As shown in \cref{fig:current}, the two solutions are present once again but do not vary significantly, even when varying the coil current by up to $175\%$. 
In other experiments (not shown), varying currents in the other coils had a similar (lack of) effect on the two solutions.


\subsubsection{Picard-based solvers and $I_p$ normalisation} \label{sec:picard}

\begin{figure}[t!]
    \centering
    \begin{subfigure}{0.99\linewidth}
        \includegraphics[width=0.99\textwidth]{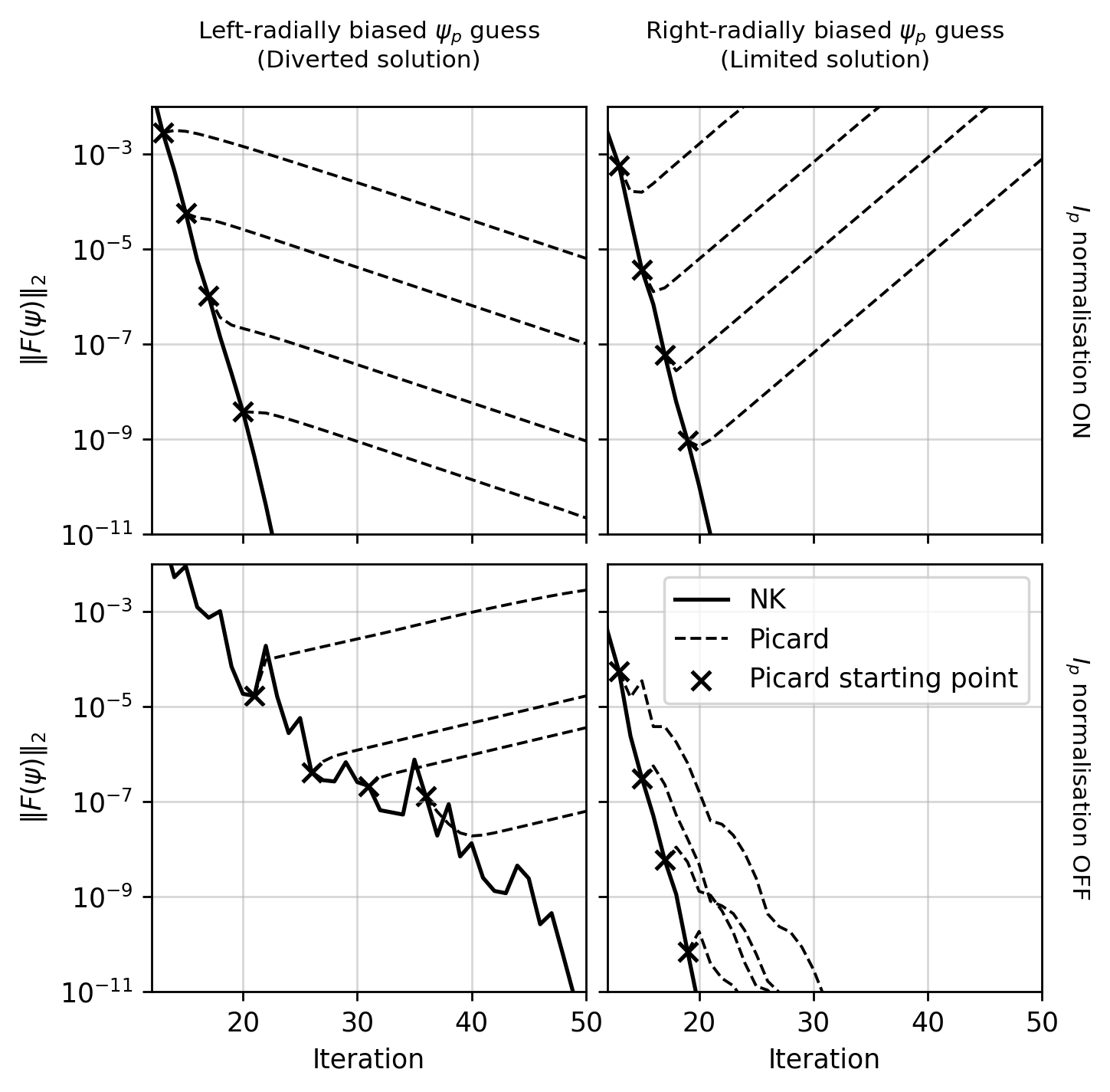}
    \end{subfigure}
    \caption{Euclidean norm of GS residual \eqref{eq:residual} against iteration number when using FreeGSNKE's NK method (solid black) or a Picard scheme (dashed black). 
    The NK simulations are initialised using a $\psi_p$ guess biased to the left (left panels) or biased to the right (right panels).
    The Picard simulations are initialised using a $\psi_p$ guess from an intermediate iteration of the NK simulation (black crosses). 
    The simulations either have $I_p$ normalisation switched on (top panels) or off (lower panels).
    }
    \label{fig:picard}
\end{figure}

Many widely used equilibrium codes rely on (possibly stabilised) Picard iterations rather than Newton-based solution schemes.
It is therefore worth carrying out a brief investigation into whether the two solution branches identified here can be detected using a Picard scheme.
Here we do not use deflated continuation and instead fix the plasma profile parameters and coil currents (as in \cref{sec:GS_problem}). 
\cref{fig:picard} displays the Euclidean norm of the residual \eqref{eq:residual} against the iteration number when using either FreeGSNKE's NK method (solid lines) or a Picard scheme (dashed lines).
For the NK method, we initialise the solver with two different guesses for $\psi_p$, each of which converges to either the diverted (left panels) or limited solution branch (right panels). 
We also explore the impact of enabling (upper panels) or disabling (lower panels) $I_p$ normalisation, a typical setting in most equilibrium codes which fixes the total plasma current---recall \eqref{eq:normalisation}.
In all four cases, we see that the NK method converges to a solution, in line with the previous results.
We use intermediate solutions provided by the NK method to initialise the Picard method (black crosses).
Given an initial guess $\psi^0$, the $k$-th Picard iteration is given by
\begin{equation*}
    \psi^{k+1}(R,Z) = \psi^k(R,Z) - \frac{1}{2} \left[ F(\psi^k(R,Z)) + F(\psi^k(R,-Z)) \right],
\end{equation*}
where we symmetrise the residual update to mitigate the typical instability of the Picard iterations.

While not an exhaustive study, \cref{fig:picard} suggests that only one of the two branches is a stable fixed point of the Picard iterations.
With $I_p$ normalisation enabled, the Picard method consistently converges to the diverted solution for a range of left-biased guesses but fails to converge to the limited solution when the initial guess is biased to the right.
Conversely, when $I_p$ normalisation is disabled, the Picard method instead converges to the limited solution and fails to find the diverted one.
This supports earlier studies by \cite{ilgisonis2004} (recall \cref{sec:intro}), which had mentioned the role of $I_p$ normalisation in Picard iterations.
This also suggests that, in practice, the impact of multiple solutions may vary across studies, depending on the equilibrium code and the nonlinear solver in use.

\section{Discussion and future work} \label{sec:discussion}

We have demonstrated that the static forward GS problem can exhibit multiple solutions when solved in a physically-relevant setup, with an integral free-boundary condition, realistic plasma current density profiles, and external conductor currents in the MAST-U tokamak. 
By utilising both FreeGSNKE and deflated continuation, our numerical experiments revealed that two distinct solution branches exist when varying parameters such as the plasma current ($I_p$), plasma current density profile coefficients ($\alpha_i$, $\beta_i$), or coil currents ($I^c_j$).
The solutions identified had significantly different shapes and positions, with one being more deeply confined (and for the most part diverted) while the other was more shallowly confined (and always limited). 

One key difference between the results presented here and in prior studies investigating the presence of multiple GS solutions is the restriction imposed by the integral boundary condition \eqref{eq:Grad--Shafranov_BC}.
Unlike boundary conditions where the solution takes a constant (but free) value on the domain boundary, \eqref{eq:Grad--Shafranov_BC} globally couples the boundary flux values with those on the domain’s interior.
Consequently, this significantly constrains the solution space for $\psi$ and not only restricts the boundary flux values but also strongly influences the internal structure of possible solutions.
This perhaps makes the emergence of even more equilibrium solutions more difficult, though we would not rule out their presence without wider study on free-boundary equilibria in other tokamaks (and perhaps using other forward equilibrium codes). 

These findings suggest that care must be taken when using forward GS solvers as they do not currently account for the presence of multiple equilibria and may converge to either of the equilibria we have seen here (depending on the inputs and the initial plasma flux guess).
As hinted in \cref{sec:picard}, only one solution branch appears to be a stable fixed point for Picard-based solvers in our experiments. 
While, on the one hand, this suggests that Picard solvers may be less prone to multiple solutions, it also further highlights that care is needed when comparing solutions between different codes, as, in the presence of multiple solutions, the adopted solver may affect which solution is identified.
In particular, while the outboard limited equilibrium may not appear to be a ``physically valid'' solution, the GS solver by itself cannot determine this. 
One way to select the appropriate equilibrium could be to ``bias'' the initial plasma flux guess more toward the inboard side (in this particular case), however, this is not systematic and is not guaranteed to return the inboard diverted solution.
A more complex but robust approach would be to incorporate some form of deflation into GS solvers to identify any possible equilibria before selecting the ``correct'' one based on experimental measurement data (e.g.~from magnetic probes or fluxloops).
We note this process may, however, become more challenging in situations when the two solutions obtained become less distinguishable (for instance, at larger $I_p$ as in \cref{fig:Ip_solutions}) and in predictive modelling when experimental data is unavailable.

Along similar lines, the presence of distinct solution branches may also affect time-dependent equilibrium calculations in which the poloidal flux is evolved alongside currents in the conducting metal structures.
These evolutive solvers play an important role in predictive ``feed-forward'' and ``feedback'' plasma scenario and control modelling.
Depending on the algorithm being used (for example, the one implemented in FreeGSNKE), each time step of the evolution may require solving tens of static GS problems. 
If any of these internal GS solver were to select the ``incorrect'' branch — or ``flicker'' back and forth between branches — it could undermine the stability/validity of the overall simulation.

Having identified multiple solution branches in MAST-U with FreeGSNKE, we think it would be of interest in the future to investigate whether or not these (and perhaps other) branches exist when using alternative equilibrium codes, tokamak geometries, and plasma conditions.
Further study will aid in the development of more robust algorithms for plasma control and scenario forecasting and help avoid convergence issues in future equilibrium calculations.




\section*{Acknowledgements}

We would like to thank the anonymous reviewers for their insightful comments on improving this manuscript. 

This work was funded by the EPSRC Energy Programme (grant number EP/W006839/1) by EPSRC grants EP/R029423/1 and EP/W026163/1, and by the Donatio Universitatis Carolinae Chair ``Mathematical modelling of multicomponent systems''.

For the purpose of open access, the authors have applied a Creative Commons Attribution (CC BY) licence to any author accepted manuscript version arising from this submission.
To obtain further information, please contact publicationsmanager@ukaea.uk.


\section*{Data availability}
The code scripts and data used in this paper can be found here: \url{https://doi.org/10.14468/bqts-4g67}.

A simple Python implementation of deflated continuation, deployed on a few polynomial and ordinary differential equation examples \href{https://github.com/kpentland/deflated_continuation_basic}{can be found on GitHub}.


\section*{Declarations}
The authors have no conflicts of interest to declare.

\begingroup
\small                        
\bibliographystyle{abbrvnat}  
\bibliography{references}  
\endgroup


\end{document}